\documentclass[twocolumn]{article}

\newcommand{\beq}{\begin{equation}}
\newcommand{\eeq}{\end{equation}}

\newcommand{\answer}{\tau = 1.168 \pm 0.007~{\rm s}}

\usepackage{graphics}

\begin{document}
%\draft

\title{Precision Measurement of the
Lifetime of the $3d\,^2\!D_{5/2}$ state in $^{40}$Ca$^+$}
\author{P. A. Barton, C. J. S. Donald, D. M. Lucas, D. A. Stevens, A. M. Steane, D. N. Stacey\\
\small Centre for Quantum Computation,
Department of Atomic and Laser Physics, University of Oxford, \\
\small Clarendon Laboratory, Parks Road, Oxford OX1 3PU, England.
}

\date{10 February, 2000}

\maketitle

\begin{abstract}
We report a measurement of the lifetime of
the $3d \,^2\!D_{5/2}$ metastable level in $^{40}$Ca$^+$,
using quantum jumps of a single cold calcium ion in a linear
Paul trap. 
The $4s \, ^2\!S_{1/2}$--$3d \,^2\!D_{5/2}$ transition
is significant for
single-ion optical frequency standards, astrophysical references,
and tests of atomic structure calculations. 
We obtain $\answer$ from observation of nearly $64,000$
quantum jumps during $\sim 32$ hours.
Our result is more precise and 
significantly larger than previous measurements. 
Experiments carried out to quantity systematic effects
included a study of
a previously unremarked source of systematic error, namely
excitation by the broad background of radiation emitted by
a semiconductor diode laser. Combining our result with atomic
structure calculations yields $1.20 \pm 0.01~{\rm s}$ for
the lifetime of $3d\,^2\! D_{3/2}$.
We also use quantum jump observations to demonstrate photon
anti-bunching, and to estimate background pressure and heating
rates in the ion trap.
\end{abstract}

%\pacs{32.70.Cs, 95.30.Ky, 42.50.Lc}
% atomic lifetimes, atomic/astro spectra, quantum jumps

\maketitle

\section{Introduction}

In this paper we present a measurement of the natural lifetime $\tau$ of
the $3d\,^2\! D_{5/2}$ metastable level in singly-ionised calcium, using quantum
jumps of a single cold calcium ion in a linear
Paul trap. The $3D_{5/2}$ level
is of interest as the source of an optical frequency standard at 729~nm
with $1/\tau \sim 1$ Hz natural linewidth \cite{93:Plumelle}, as a
means of testing atomic structure calculations
\cite{88:Ali,90:Zeippen,91:Guet,92:Vaeck,93:Brage,95:Liaw,96:Biemont},
and as a diagnostic in
astrophysics \cite{Bk:Osterbrock,90:Zeippen}. In addition, our experiments
offer insights into
the diagnostics on the performance of the Paul trap, and highlight the need
to take into account the spectral properties of semiconductor diode laser
emission when such devices are used in atomic physics experiments.

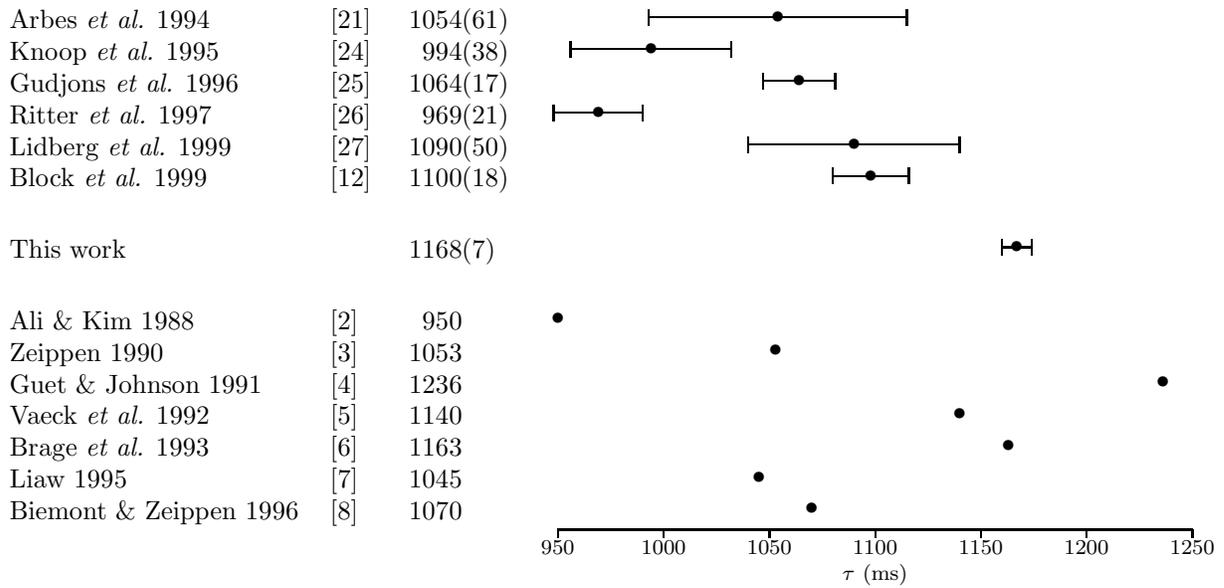
\begin{figure}[tp]
\begin{center}
  \begin{tabular}{ll@{\hspace{5mm}}r@{\hspace{0cm}}l@{\hspace{2cm}}c}
Arbes {\em et al.} 1994 & \protect\cite{94:Arbes} & 1054 & (61) & 
\begin{picture}(160,8)
\put(43.20,4.00){\makebox(0,0){$\bullet$}}
\put(-5.60,1.00){\line(0,1){6.00}}
\put(-5.60,4.00){\line(1,0){97.60}}
\put(92.00,1.00){\line(0,1){6.00}}
\end{picture} \\
Knoop {\em et al.} 1995 & \protect\cite{95:Knoop} & 994 & (38) & 
\begin{picture}(160,8)
\put(-4.80,4.00){\makebox(0,0){$\bullet$}}
\put(-35.20,1.00){\line(0,1){6.00}}
\put(-35.20,4.00){\line(1,0){60.80}}
\put(25.60,1.00){\line(0,1){6.00}}
\end{picture} \\
Gudjons {\em et al.} 1996 & \protect\cite{96:Gudjons} & 1064 & (17) & 
\begin{picture}(160,8)
\put(51.20,4.00){\makebox(0,0){$\bullet$}}
\put(37.60,1.00){\line(0,1){6.00}}
\put(37.60,4.00){\line(1,0){27.20}}
\put(64.80,1.00){\line(0,1){6.00}}
\end{picture} \\
Ritter {\em et al.} 1997 & \protect\cite{97:Ritter} & 969 & (21) & 
\begin{picture}(160,8)
\put(-24.80,4.00){\makebox(0,0){$\bullet$}}
\put(-41.60,1.00){\line(0,1){6.00}}
\put(-41.60,4.00){\line(1,0){33.60}}
\put(-8.00,1.00){\line(0,1){6.00}}
\end{picture} \\
Lidberg {\em et al.} 1999 & \protect\cite{99:Lidberg} & 1090 & (50) & 
\begin{picture}(160,8)
\put(72.00,4.00){\makebox(0,0){$\bullet$}}
\put(32.00,1.00){\line(0,1){6.00}}
\put(32.00,4.00){\line(1,0){80.00}}
\put(112.00,1.00){\line(0,1){6.00}}
\end{picture} \\
Block {\em et al.} 1999 & \protect\cite{99:Block} & 1100 & (18) & 
\begin{picture}(160,8)
\put(80.00,4.00){\makebox(0,0){$\bullet$}}
\put(65.60,1.00){\line(0,1){6.00}}
\put(65.60,4.00){\line(1,0){28.80}}
\put(94.40,1.00){\line(0,1){6.00}}
\end{picture} \vspace*{1.5em}\\
This work &  & 1168 & (7) & 
\begin{picture}(160,8)
\put(135.20,4.00){\makebox(0,0){$\bullet$}}
\put(129.60,1.00){\line(0,1){6.00}}
\put(129.60,4.00){\line(1,0){11.20}}
\put(140.80,1.00){\line(0,1){6.00}}
\end{picture} \vspace*{1.5em}\\
Ali \& Kim 1988 & \protect\cite{88:Ali} & 950 & & 
\begin{picture}(160,8)
\put(-40.00,4.00){\makebox(0,0){$\bullet$}}
\end{picture} \\
Zeippen 1990 & \protect\cite{90:Zeippen} & 1053 & & 
\begin{picture}(160,8)
\put(42.40,4.00){\makebox(0,0){$\bullet$}}
\end{picture} \\
Guet \& Johnson 1991 & \protect\cite{91:Guet} & 1236 & & 
\begin{picture}(160,8)
\put(188.80,4.00){\makebox(0,0){$\bullet$}}
\end{picture} \\
Vaeck {\em et al.} 1992 & \protect\cite{92:Vaeck} & 1140 & & 
\begin{picture}(160,8)
\put(112.00,4.00){\makebox(0,0){$\bullet$}}
\end{picture} \\
Brage {\em et al.} 1993 & \protect\cite{93:Brage} & 1163 & & 
\begin{picture}(160,8)
\put(130.40,4.00){\makebox(0,0){$\bullet$}}
\end{picture} \\
Liaw 1995 & \protect\cite{95:Liaw} & 1045 & & 
\begin{picture}(160,8)
\put(36.00,4.00){\makebox(0,0){$\bullet$}}
\end{picture} \\
Biemont \& Zeippen 1996 & \protect\cite{96:Biemont} & 1070 & & 
\begin{picture}(160,8)
\put(56.00,4.00){\makebox(0,0){$\bullet$}}
\end{picture} \\
 & & & & {\footnotesize 
\begin{picture}(160,8)
\put(-40,8){\line(1,0){240}}
\put(-40.00,8){\line(0,-1){3}}\put(-40.00,4){\makebox(0,0)[tc]{$950$}}
\put(0.00,8){\line(0,-1){3}}\put(0.00,4){\makebox(0,0)[tc]{$1000$}}
\put(40.00,8){\line(0,-1){3}}\put(40.00,4){\makebox(0,0)[tc]{$1050$}}
\put(80.00,8){\line(0,-1){3}}\put(80.00,4){\makebox(0,0)[tc]{$1100$}}
\put(120.00,8){\line(0,-1){3}}\put(120.00,4){\makebox(0,0)[tc]{$1150$}}
\put(160.00,8){\line(0,-1){3}}\put(160.00,4){\makebox(0,0)[tc]{$1200$}}
\put(200.00,8){\line(0,-1){3}}\put(200.00,4){\makebox(0,0)[tc]{$1250$}}
\put(80.00,-5){\makebox(0,0)[tc]{$\tau$ (ms)}}
\end{picture}} \\
  \end{tabular}\par
\end{center}
\caption{Values of $\tau$ experimentally measured (points with error bars) and
theoretically calculated {\em ab initio} (circles). The first two columns give
the reference,
the third gives $\tau$ in milliseconds. Earlier measured values are not shown since
they have considerably larger experimental uncertainty.}
\label{f_tau}
\end{figure}

Our result is $\answer$, and is shown together with other
recent measurements and theoretical predictions in figure \ref{f_tau}.
Our result is higher (a longer lifetime) than all the previous 
measurements, and differs by several standard deviations from most of them. 
It is possible that this discrepancy is at least partly due to a previously
unrecognised source of 
systematic error, namely the presence of light of wavelength in the 
vicinity of 854~nm in the beam produced by a semiconductor diode laser emitting 
predominantly at 866~nm. This is discussed in section \ref{s_854}.

Precise knowledge of atomic structure for atoms or ions with a single electron
outside closed shells is currently in demand for the analysis of
atomic physics tests of electro-weak theory, especially measurements of parity
violation in Cs \cite{99:Bennett}. There has been a long-standing discrepancy at the 2\%
level between measured and theoretically predicted rates for electric dipole
transitions \cite{93:Jin}. Up until now measurements of the metastable lifetimes
(electric quadrupole transition rates) in Ca$^+$ 
have not been sufficiently precise to provide an independent test of
{\em ab initio}
calculations at this level of precision. Our measurement precision is
$0.6$\,\%. Furthermore the electric quadrupole transition rate is harder
to calculate accurately (recent calculated values have a 15\% spread), so our result
is of particular interest to atomic structure theory. 

The paper is organized as follows. First we discuss, in section \ref{s_method}, 
the central features of our experimental method to measure $\tau$. Details
of the apparatus are provided in section \ref{s_app}. This is a 
completely new apparatus which has not been described elsewhere, so 
we give a reasonably full description.

Section \ref{s_sys} presents our study of systematic effects in the
experiment. These include 
collisions with the background gas, off-resonant 
excitation of the 854~nm $3D_{5/2}$--$4P_{3/2}$ transition and of 
the 850~nm $3D_{3/2}$--$4P_{1/2}$ transition, heating of the 
trapped ion, and noise in the fluorescence signal.
Section \ref{s_anti} contains a clean demonstration of photon
anti-bunching, using the random telegraph method,
and section \ref{s_tau} presents the final accurate
measurements of $\tau$. 

\section{Experimental method}  \label{s_method}

\begin{figure}[h]
\centerline{\resizebox{10cm}{!}{\includegraphics{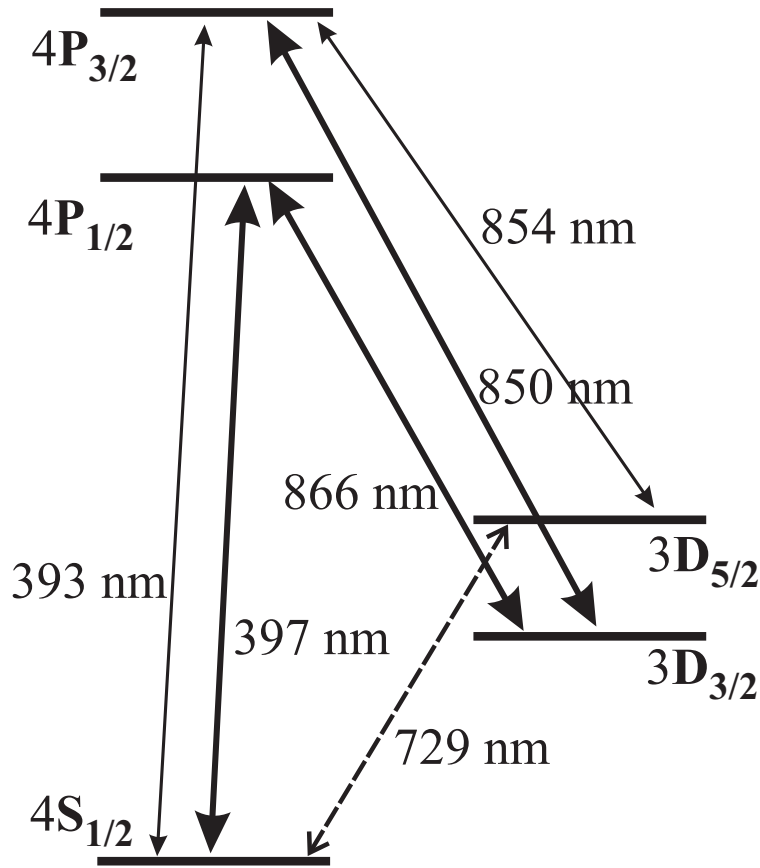}}}
\rule{0pt}{24 pt} 
\caption{Low-lying energy levels of $^{40}$Ca$^+$ and associated transitions.
Lasers at 397 nm, 866 nm and 850 nm drive the corresponding
transitions in the experiments. The other transitions are significant to the
interpretation of the measured signals and to systematic effects.}
\label{f_levels}
\end{figure}

\begin{figure}[h]
\centerline{\resizebox{14cm}{!}{\includegraphics{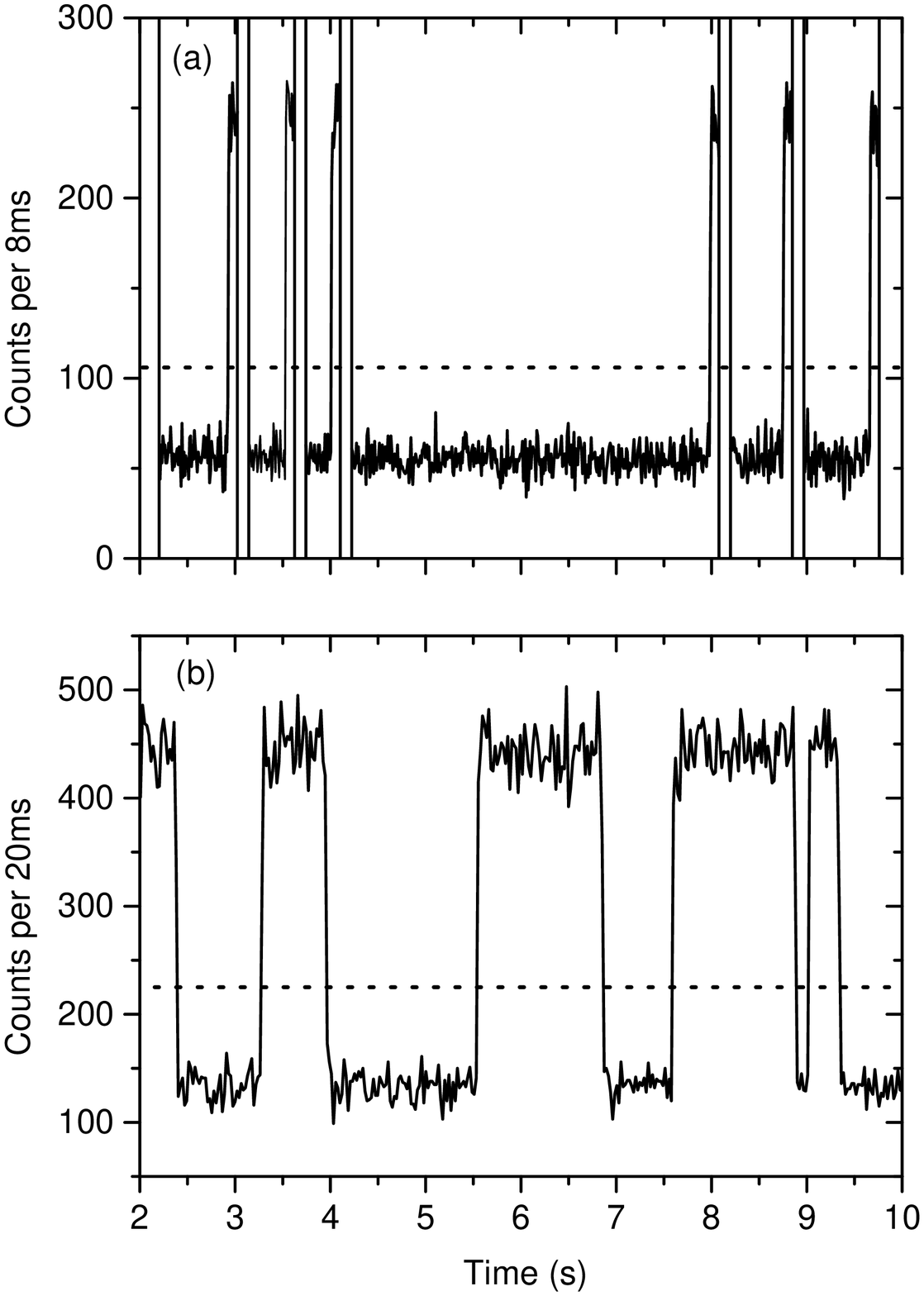}}}
\rule{0pt}{24 pt}
\caption{Observed fluorescence signals. The vertical axis is the number
of counts given by the photomultiplier during one counting bin, the
horizontal axis is time. The horizontal dashed lines show the threshold
settings for the data analysis.  (a) Shutter method, in which 850 nm laser is blocked
during dark periods. 
The vertical lines indicate shutter opening and closing events.
(b) Random telegraph method, in which 850 nm laser
is permanently on. (a) was used for accurate $\tau$ measurements, (b) for
$g^{(2)}$ measurements and various systematic studies.}
\label{f_sig}
\end{figure}

\begin{figure}[h]
\centerline{\resizebox{15cm}{!}{\includegraphics{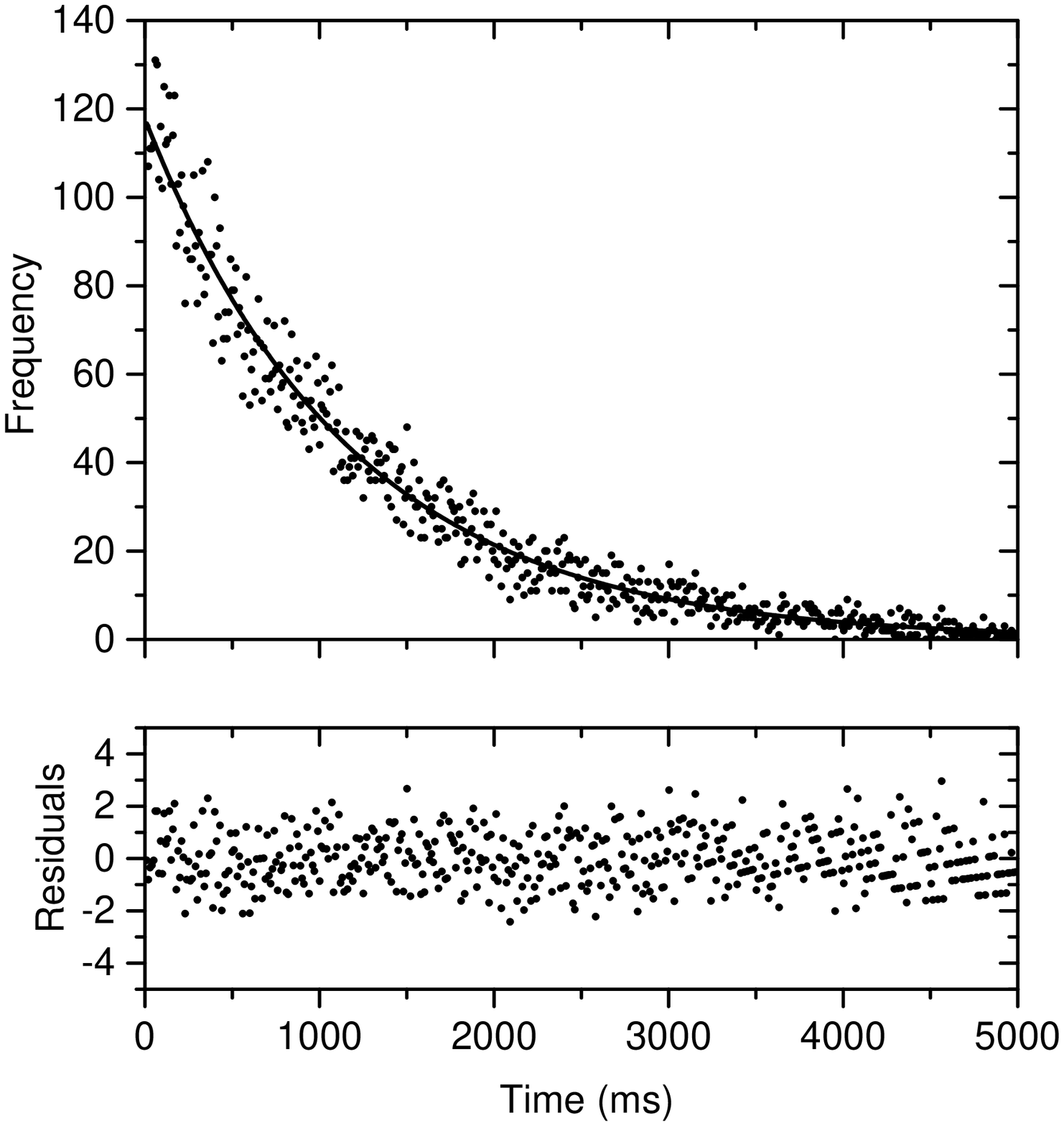}}}
\rule{0pt}{24 pt}
\caption{Typical data set from a single 8-hour run. The figure shows
a histogram of the measured dark period durations $x_i$, and the fitted
exponential curve produced by our analysis procedure (see text).
The residuals are shown on an expanded scale, in the form
(data$-$fit)/$\surd$(fit).
In this example the analysis
gave $A=117.7 \pm 1.5,\;\gamma = 0.853 \protect\pm 0.009$.
}
\label{f_hist}
\end{figure}

The experimental method was identical, in 
principle, to that adopted by Block {\em et al.} \cite{99:Block};
preliminary experiments were carried out somewhat differently (see
section \ref{s_anti}). A single ion of 
$^{40}$Ca$^+$ is trapped and laser-cooled to around 1 mK. The transitions of
interest are shown in figure \ref{f_levels}.
Laser beams at 397~nm and 866~nm 
continuously illuminate the ion, and the fluorescence 
at 397~nm is detected by a photomultiplier.
The photon count signal is accumulated for
periods of duration $t_b = 10.01$~ms 
(of which 2.002~ms is dead time), and logged. In our studies of systematic
effects, and for our demonstration of photon antibunching, $t_b$ was set at $22.022$ ms.
A laser at 850~nm drives the $3D_{3/2}$--$4P_{3/2}$ transition. 
The most probable decay route from $4P_{3/2}$ is to the $4S_{1/2}$ ground state;
alternatively, the ion can return to $3 D_{3/2}$. However, about 1 decay in 18
occurs to $3 D_{5/2}$, the metastable ``shelving" level of interest.
At this point the fluorescence abruptly disappears and the observed 
photon count signal falls to a background level. 
A shutter on the 850~nm laser beam remains open for 100 ms before it is closed,
which gives ample time for shelving.
Between 5 and 10~ms after the shutter has closed,
we begin one ``observation'', i.e., we start to
record the photomultiplier count signal (see figure \ref{f_sig}a). We keep 
observing the photon count,
in the 10~ms bins, until it abruptly increases to a 
level above a threshold. This is set between the background 
level and the level observed when the ion fluoresces continuously. The signature
for the end of a dark period is taken to be ten consecutive bins above threshold.
We record the number of 10~ms bins in the observed `dark' period.
The 100 ms period of fluoresence also serves to
allow the 397~nm and 866~nm lasers to cool the ion. After this
we re-open the shutter on the 850~nm laser. 
This process is repeated for long periods of time (1 to 8 
hours), the laser intensities being also monitored and the frequencies
servo-controlled. 
Subsequent analysis of the large collection of dark times 
consists primarily of gathering them into a histogram, and fitting the expected 
exponential distribution, in order to derive the decay rate from the shelved 
state (see figure \ref{f_hist}).

Note that we do not measure the length of the dark period from when the 
ion is first shelved. This is not necessary, since the probability of
decay is independent of how long the ion has been in the 
metastable state. This gives us time to block the 850~nm light, in order
to prevent both off-resonant excitation of the ion by this light, and the 
possibility of missed quantum jumps should this light rapidly (in less than 5 
ms) re-shelve a decayed ion. 

The data from a given run were analysed as follows. The raw data
consist of a series of counts indicating the average fluorescence level 
in each bin of duration $t_b$. A threshold is
set, typically at $(2 S_{\rm dark} + S_{\rm bright})/3$, where
$S_{\rm dark}$ is the mean count observed during dark periods,
and $S_{\rm bright}$ the mean count observed during 
fluorescing periods. This setting is chosen
because bright periods have more noise than dark periods.
The number of consecutive bins below threshold is a single
dark-time measurement $x_i$. The $x_i$ are expected to be distributed
according to the exponential decay law, with 
Poissonian statistics describing the departures from the mean.
It is appropriate to use a Poissonian fitting
method, rather than least squares, because of the small numbers
involved in part of the distribution (at large $t$).
If $n(t_i)$ is the number of $x_i$
equal to $t_i/t_b$ then the cost function is
  \begin{eqnarray}
- \ln \left[ \prod_{i=0}^m P\left( n(t_i) \right) \right]
&=& \sum_{i=0}^m \left[ A e^{-\gamma t_i} + \ln \left( n(t_i)! \right) 
 \right.   \nonumber \\
& & \left. - n(t_i) \ln A + n(t_i) \gamma t_i  \right]
  \end{eqnarray}
where $m$ is the number of bins, and $A$ and $\gamma$ are
two fitted parameters (obtained by minimising the cost function);
they are the amplitude and decay rate
in the assumed exponential decay $A \exp(-\gamma t)$
of population of $D_{5/2}$. 
The residuals shown in figure \ref{f_hist} indicate that our
data are well fitted by an exponential function.
Only dark times of duration less than 5~s are included in
the fit, since our data collection procedure misses some dark times
longer than this.
We found that the statistical error in the fitted parameters
was consistent with
the expected $\sqrt{N}/N$ value, where $N$ is the number
of $x_i$ in the whole data set.

\section{Apparatus}  \label{s_app}

\begin{figure}[h]
\centerline{\resizebox{8cm}{!}{\includegraphics{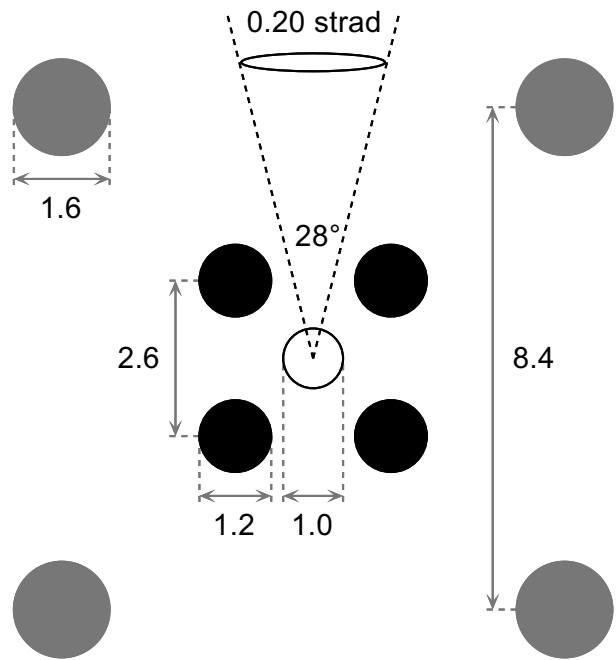}}}
\rule{0pt}{24 pt}
\caption{End view of the linear ion trap electrode arrangement, to scale, showing
the a.c. trap electrodes (solid black circles) and d.c. compenstation electrodes
(grey circles). Dimensions are in mm. The position of the d.c. endcaps is also 
indicated (open circle); the separation between the two opposing endcaps is $7.2$ mm.
The solid angle in which fluorescence is collected by the imaging system is $0.2$
steradians. 
}
\label{f_electrodes}
\end{figure}

We use a linear radio-frequency (r.f.) Paul trap, combined with an
all-diode laser system, to isolate and cool a single ion of
$^{40}$Ca$^+$. The electrodes are made from stainless steel rods
and mounted on two supports made from machinable ceramic (Macor); an end view
is shown in figure~\ref{f_electrodes}. The radial r.f.\ electrodes,
of diameter 1.2\,mm, are centred at the corners of a square of side
$2.6$\,mm.  The d.c.\ endcap electrodes, of diameter 1.0\,mm are
centred on the $z$-axis and positioned 7.2\,mm apart. In addition to
the electrodes which comprise the trap, there are a further four
1.6\,mm diameter electrodes positioned in a similar configuration to
that of the r.f.\ electrodes, but centred at the corners of a
8.4\,mm square. These electrodes allow potentials to be added to
compensate for stray electric fields in the trapping region.

\begin{figure}[h]
\centerline{\resizebox{17cm}{!}{\includegraphics{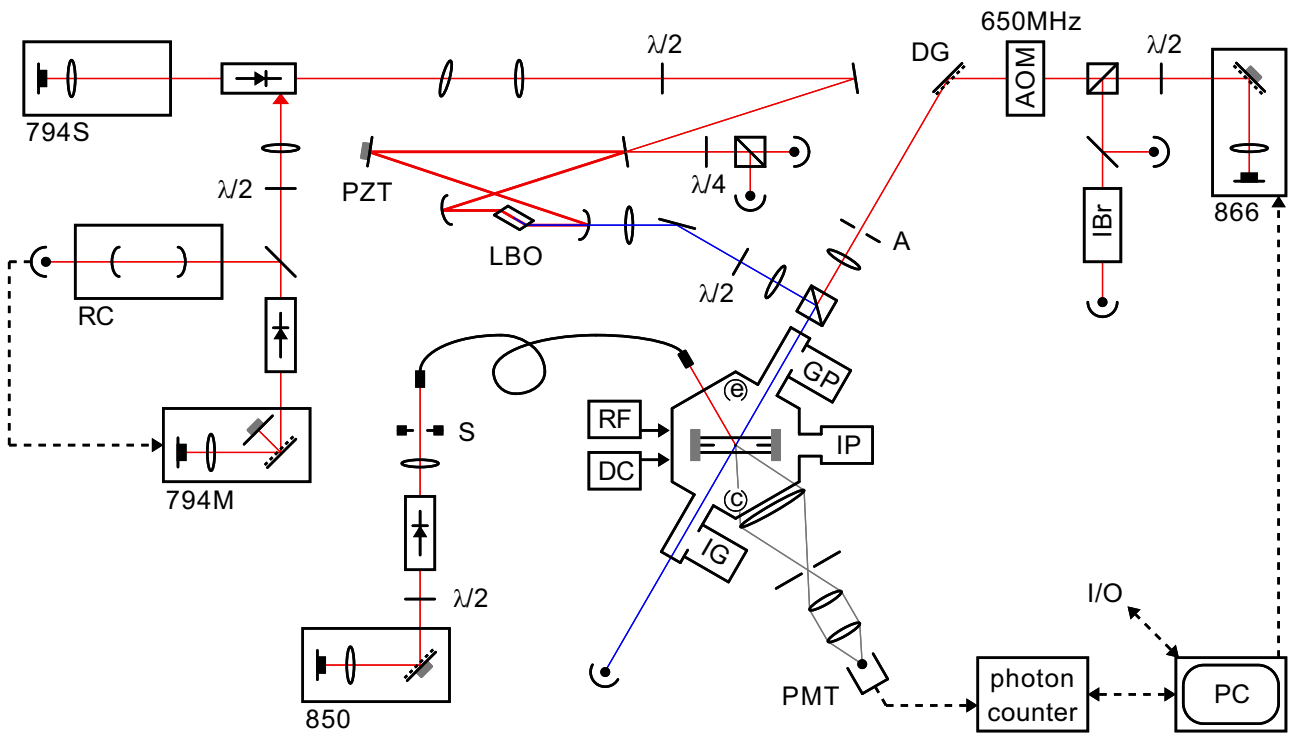}}}
\rule{0pt}{24 pt}
\caption{Main features
of the optical set-up; see section \protect\ref{s_app} for key.}
\label{f_exp}
\end{figure}

The complete experimental apparatus is shown schematically in
figure~\ref{f_exp}, to which we refer in the remainder of this
section. The trap electrodes lie at the centre of a hexagonal
stainless steel vacuum chamber. A high-voltage a.c.\ source {\sf RF}
supplies a drive voltage of frequency $6.2$\,MHz and peak-to-peak
amplitude 135\,V for the radial electrodes, while high-voltage d.c.\
supplies {\sf DC} provide the voltages for the endcaps (95\,V in the
present experiments) and compensation electrodes (typically around
60\,V are applied to the upper two compensation electrodes, while the
lower two are grounded).

The central chamber is pumped by a 25\,l/s ion pump {\sf IP} and a 30\,l/s
getter pump {\sf GP}. An ion gauge {\sf IG} on the opposite side of the
main chamber monitors the pressure; this is below $2\times
10^{-11}$\,torr, the limit of the gauge's sensitivity. To produce calcium
ions in the trapping region, we use a calcium oven {\sf c} and electron
gun {\sf e}. The oven is a thin-walled stainless steel tube
filled with calcium granules, closed by crimping at each end and with a small
hole at the centre pointing towards the trap region. The oven is
heated by passing a 6\,A current along its length, which produces a beam
of calcium atoms. The electron gun consists of a tungsten filament, also
resistively heated, enclosed within a grounded stainless steel ``grid'' with
respect to which it is negatively biased by 50\,V. To load the trap, the
oven and electron gun are heated for a few minutes, the latter being
left on for 10\,s after the oven has been turned off. We capture a
small cloud of approximately 10 calcium ions using this procedure.
This cloud is reduced to a single ion by applying to one of the
endcaps a low-voltage ``tickle'' oscillation, close to the axial
resonance frequency of the trap, expelling ions until only one
remains in the trap.

Violet light at 397\,nm is generated by frequency-doubling 794\,nm
light from a master-slave diode laser system. A grating-stabilized
master diode {\sf 794M}, with a linewidth below 1\,MHz, is locked to
a stabilized low-finesse reference cavity {\sf RC} and used to inject
a slave diode {\sf 794S}. Light from the slave is frequency-doubled
by a 10\,mm long Brewster-cut lithium triborate crystal {\sf LBO} in
an external enhancement cavity; H{\"a}nsch-Couillaud polarization
analysis of light reflected by the cavity provides a feedback signal
for a piezo-mounted mirror {\sf PZT} used to lock the cavity length
to the fundamental light. The measured characteristics of the
doubling cavity are: finesse 130, enhancement 43, mode-matching
efficiency 94\%, input power 90\,mW at 794\,nm, output power 0.50\,mW
at 397\,nm. Correcting for losses at the exit face of the crystal and
the output coupler gives an internal crystal efficiency $\gamma =
54(5)\,\mu$W/W$^2$, some 20\% greater than previously reported
values~\cite{93:Bourzeix,92:Adams} and about 75\% of
the theoretical optimum
efficiency calculated using a Boyd-Kleinman analysis~\cite{68:Boyd} and
an effective non-linear coefficient for LBO of $d_{\mbox{eff}} =
0.855$\,pm/V~\cite{Casix}. The 397\,nm beam passes through a
$\lambda/2$ waveplate and polarizing beam splitter (PBS)
cube to provide intensity
control, and a lens focuses it to a spot size of $200\times 30\,\mu$m
(measured with a CCD camera) at the centre of the trap. The beam
power used is typically 0.2\,mW.

A grating-stabilized diode laser {\sf 866} provides repumping light
at 866\,nm whose intensity is adjusted by a $\lambda/2$ waveplate and
PBS cube. A heated iodine bromide vapour cell {\sf IBr} provides an
absolute frequency reference---unfortunately not suitable for
reliable locking---and a triple-pass acousto-optic modulator {\sf
AOM} shifts the laser light some 650\,MHz into resonance with the
$3 D_{3/2} \rightarrow 4 P_{1/2}$ calcium transition.
Spontaneous light near 854\,nm emitted by the laser diode (see
section~\ref{s_854L}) is rejected by a diffraction grating {\sf
DG} and aperture {\sf A}. The transmitted 866\,nm light is superimposed on
the 397\,nm beam using a PBS
cube and focused onto the trap. The spot
size used in the final lifetime measurements was $250\times
130\,\mu$m; the maximum beam power was 2\,mW.

The light at 850\,nm to shelve the ion is provided by a third grating-stabilized
laser diode {\sf 850} situated on a separate optical table,
and is directed into the ion trap via a
polarization-preserving monomode optical fibre. The intensity is
controlled by a $\lambda/2$ waveplate before the optical isolator. A
mechanical shutter {\sf S} before the fibre
allows complete extinction of this
light. The spot size at the trap is $450\times 450\,\mu$m and the
maximum power $0.5$\,mW.

For clarity in figure~\ref{f_exp} the detection optics are
shown in the plane of the diagram: they are actually vertically above
the ion trap. A wide-aperture compound lens gathers fluorescence
emitted by the trapped ion and images it onto an aperture to reject
scattered light; further lenses re-image the light, via a violet
filter, onto a photomultiplier {\sf PMT} connected to a gated photon
counter. The net collection efficiency of the detection system,
including the 16\% quantum efficiency of the PMT at 397\,nm, is
approximately $0.12$\%. The peak photon count rate above background
for a single cooled ion is typically 32\,kHz.

A personal computer {\sf PC} is used for data acquisition and control
of the experiment; in particular it provides timing, logs PMT count
data, controls the shutter {\sf S}, and eliminates long-term drift of
the 866\,nm laser by locking it to the fluorescence signal from the
trapped ion at the end of each 20\,s acquisition period.

\section{Searches for systematic effects} \label{s_sys}

We now consider effects which alter the measured shelving periods from those
appropriate to an unperturbed ion subject only to spontaneous decay.  These are of two
distinct types: those which alter the shelving periods themselves, because the ion is perturbed
in some way, and those which cause systematic error in the process of measurement.  We
consider the two in turn.  The significant difference between our final result and previous
work calls for a detailed discussion here.

\subsection{Perturbations to the ion} \label{s_854}

       When the ion is fluorescing, it is cycling between the levels $4\,^2 S_{1/2},\;4\,^2 P_{1/2}$
and $3\,^2 D_{3/2}$. 
We denote this system of levels by $\Lambda$.  During the shelving period, the ion is subject to
radiation from two of the lasers (397 nm and 866 nm), to
thermal radiation, and to the fields associated with the trap.  There may also be collisions
with the background gas.  Any of these perturbations can transfer the ion to the $\Lambda$ system. 
We consider them in turn.

\subsubsection{Electric field}

The ion experiences a static electric field because of imperfect compensation.
The most significant effect on the internal state of the ion 
is to mix the $3D_{5/2}$ and $4P_{3/2}$ levels, so that the measured
lifetime is shortened by the presence of the induced $4P_{3/2}$--$4S_{1/2}$ strong
electric dipole transition.  In fact, using
the known matrix elements, one finds that the effect is negligible; the induced transition
probability is $9.0 \times 10^{-14}E^2$ s$^{-1}$, so that a field of 300 kV m$^{-1}$
(three orders of magnitude larger
than the typical compensating fields used) would be necessary to produce a 1\% reduction in
the lifetime.  We note, however, that there is another effect associated with an imperfectly
compensated field, that of heating during the shelving period; we consider this in
section \ref{s_heat}.

\subsubsection{Laser radiation}  \label{s_854L}

From the standard theory of atom-light interaction, one finds that transitions from a
lower level (1) to a higher level (2) in the multi-level ion can be stimulated
by radiation of intensity $I$ and angular frequency $\omega_L$
at a rate $R_{12}$ (averaged over all Zeeman components)
given by
\beq
  R_{12} = \frac{2J_2+1}{2J_1 + 1} \frac{ \pi^2 c^3 }
  {\hbar \omega_{12}^3} A_{21} \frac{I}{c} g(\omega_L-\omega_{12})  \label{Rav}
\eeq
where $J_1$ and $J_2$ are the total angular momenta of the levels,
$\omega_{12}$ is the atomic resonance angular frequency, 
and $g(\omega_L-\omega_{12})$ is the normalised 
lineshape function. In our case we may assume 
Lorentzian lineshapes, so that
\beq
 g(\omega_L-\omega_{12}) = \frac{\Gamma/(2 \pi)}{(\omega_L-\omega_{12})^2 + \Gamma^2/4}
\eeq
with the linewidth $\Gamma = 1/\tau_2$ determined by the sum of all the
decay processes from the upper level.

Since Block {\em et al.} \cite{99:Block} reported a
significant dependence of the shelving time on the intensity of the repumping laser, we first
consider excitation from $3D_{5/2}$ to $4P_{3/2}$ by light at 866 nm. 
The lifetime $\tau_2$ of the $4P_{3/2}$ level has been measured to be
$6.924 \pm 0.019$ ns \cite{93:Jin}.
We use the value $A_{21} = (7.7 \pm 0.3) \times 10^6$~s$^{-1}$
from {\em ab initio} \cite{95:Liaw} and semiempirical \cite{89:Theodosiou}
atomic structure calculations. 
The $4$\% uncertainty is our own estimate, based on the variation among
the published calculations, and on the fact that
the calculations produce other electric dipole matrix elements
in agreement with experiment to better than this level of precision.
In any case a 10\% error in the value of
$A_{21}$ would have a negligible influence on our final result.

To excite the $3D_{5/2}$--$4P_{3/2}$ transition on
resonance would require radiation of wavelength 854 nm.  From equation (\ref{Rav}), we
find that with 866 nm light the
rate is $9.9 \times 10^{-5}$~s$^{-1}/$mW~mm$^{-2}$. 
The probability that the ion will subsequently decay
back to $3D_{5/2}$ is the branching
ratio $b = A_{21} \tau_2 = 0.053$, so the rate at which it will be transferred to
the $\Lambda$ system is $(1-b)R_{12} = 9.4 \times 10^{-5}$ s$^{-1}/$mW mm$^{-2}$. 
In our experiments, it is convenient to choose the
866 nm intensity such that the $3D_{3/2} \rightarrow 4P_{1/2}$ transition is saturated.  This
requires $I_{866} \gg 0.08$ mW mm$^{-2}$ if the laser
frequency is set on resonance (taking this $A_{21}$ coefficient $8.4 \times 10^6$~s$^{-1}$
\cite{95:Liaw,89:Theodosiou}),
but to avoid the need to control
this frequency precisely a much higher intensity was used, typically $1.5$ mW in a spot size
of $250 \times 130\;\mu$m, or 30 mW mm$^{-2}$. 
This results in a contribution to the depopulation rate of the
$3D_{5/2}$ level of around 0.3\% of that due to spontaneous emission to the ground level.

This is far smaller than the experimental result of Block {\em et al.} However, in our own
preliminary work we also found that there was a significant dependence of the shelving time
on the intensity of the repumper laser, of the same order of magnitude as that reported by
Block {\em et al.}, and some 200 times larger than the theoretical value given above.  This
suggested that the 866 nm laser was emitting some radiation much closer in wavelength to
854 nm which was primarily responsible for the shortening of the apparent lifetime of the
$3D_{5/2}$ level.  This laser is a semiconductor diode device, operated with an extended cavity by
use of a Littrow-mounted diffraction grating.  Without the grating, the laser would operate
close to 854 nm.  Since any laser produces spontaneous emission over its gain profile, as
well as stimulated emission at the lasing wavelength, there was in our preliminary work
radiation incident on the ion in the vicinity of 854 nm.  One might expect the intensity of this
radiation to be greatly reduced because of the long (3m) beam path.  In fact, this is not the
case, because the emitting region in the laser is small (dimensions of order microns) so the
spontaneous component is well collimated; like the coherent component, it is transported to
the trap with little loss.  

       We therefore investigated the spectrum of the light from the laser by means of a
diffraction grating, and found it to contain a broad background from 840 nm to 870 nm. 
When the total laser power near the ion trap was 2 mW, the power in the broad
component was 8 $\mu$W, and the spot sizes were similar.  This implies a mean spectral density
around 854 nm sufficient to cause de-shelving rates considerably higher than those we
observed; we ascribe the lower observed rates to the structure of the background.  This was
not resolved in our investigation, but is expected on the basis of other experimental work
\cite{85:Camparo,85:Labachelerie}
to include a long series of spikes separated by the longitudinal mode spacing of the diode
(50 GHz). 
These occur because the gain is so high in these devices that there is some amplification right
across the gain profile; evidence that this occurs in our laser is provided by the high degree
of polarization (90\%) of the background.  It is thus reasonable to conclude that
the $3D_{5/2}$--$4P_{3/2}$ atomic resonance falls between peaks in the 50 GHz spaced comb.

\begin{figure}[h]
\centerline{\resizebox{14cm}{!}{\includegraphics{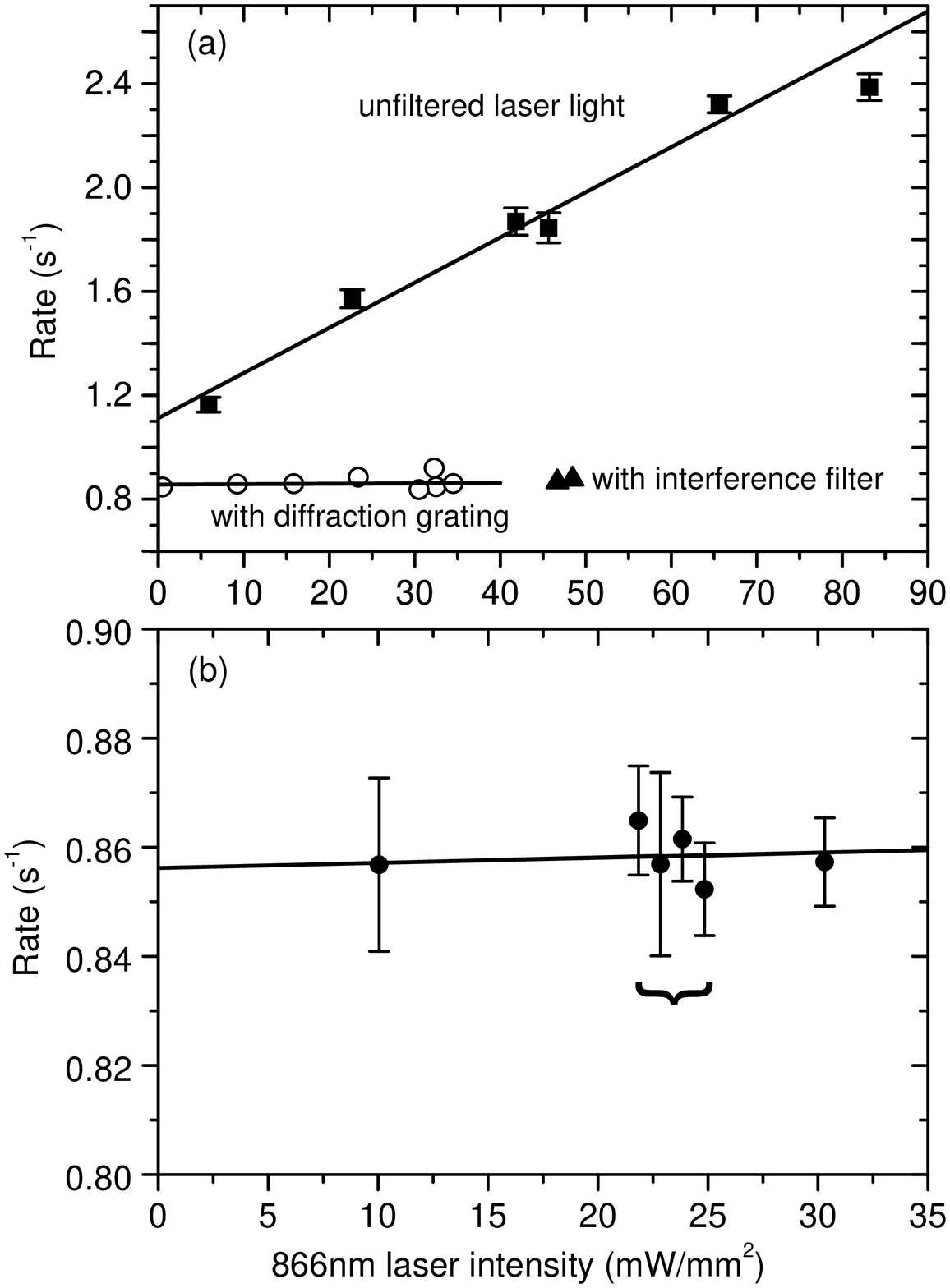}}}
\rule{0pt}{24 pt}
\caption{Measured de-shelving rates. (a) Measurements using
1 to 2 hour runs of the random telegraph method,
at various 866 nm beam intensities, to test for
intensity dependence of the rate. The lines are 2-parameter
straight line fits
to the data, with fitted intercept and slope. With the diffraction
grating and aperture in place, the slope is
consistent with zero and with the small theoretically expected
value (see text).
(b) Final data set using the shutter method. Points
taken at the same intensity are shown horizontally
offset for clarity. The error bars
are the statistical uncertainties emerging from the analysis
of each run.
The line is a single-parameter least-squares fit; it is given the theoretically
expected slope, and the best fit intercept is obtained.}
\label{f_res}
\end{figure}

       To study the effects of this background radiation we reduced its intensity at the
position of the ion.  This was done in two stages: first, we reduced it by a factor of 25 using
an interference filter centred on 866 nm.  The observed de-shelving rate fell
from $1.85 \pm 0.06$ s$^{-1}$ to $0.87 \pm 0.02$ s$^{-1}$, the two measurements
being taken at the same 866 nm intensity. 
This provided convincing evidence that the de-shelving was indeed caused by the postulated
mechanism.  We therefore replaced the filter with a diffraction grating and iris aperture,
arranged to pass light at 866 nm.  The power transmitted by the system in the vicinity
of 854 nm was then reduced compared with the unfiltered laser by three and a half
orders of magnitude (for a given intensity at 866 nm), and the remaining
light was scattered, thus increasing the
illuminated area in the vicinity of the ion by a factor measured to be 40,
making a net intensity reduction $8 \times 10^{-6}$.
Under these conditions transitions stimulated by
this radiation become much less probable than those due to the 866 nm light itself.  The
observed de-shelving rate was $0.858 \pm 0.007$ s$^{-1}$.  De-shelving rates $\gamma$ at
various settings of
the intensity $I$ of the 866 nm laser beam as measured in these various experiments are shown
in figure \ref{f_res}a.  A straight line of the form $\gamma = \gamma_0 + \alpha I$
fitted to the points for which the grating and
iris system were in place gives $\gamma_0 = 0.857 \pm 0.016$ s$^{-1}$,
$\alpha = (1.5 \pm 6) \times 10^{-3}$ s$^{-1}$/mW~mm$^{-2}$.  The
value of $\gamma_0$ is consistent with our final more accurate measurements, given in
section \ref{s_tau} below, while the slope is consistent
with zero (and with the very small theoretical value for
de-shelving by 866 nm radiation given above).
The intercept with unfiltered light is greater than that given by our final data
because our method of varying the laser intensity did not alter the unpolarized
component of the background radiation.

       Spontaneous emission from the repumper laser appears also to be likely to account
for the observations of Block {\em et al.}  It is possible that it was present in previous work on the
$3D_{5/2}$ lifetime, and unaccounted for, explaining the lower values obtained by all earlier
workers.  This does not immediately apply to Block {\em et al.}, however, because their
measurement involved an extrapolation to zero laser intensity using neutral
density filters.

The 866 nm laser might also generate radiation near 854 nm if it went multimode owing
to a degradation of the alignment of its own external cavity. If intermittent
and at a low level, this effect could occur undetected.
However, multimode operation was
found in practice to have an all-or-nothing character: if it occurred at all then
it was obvious from a large increase in the noise of the fluorescence signals, and
in such a case the data were discarded.

       The grating and iris system were in place for our final data sets. 
Figure \ref{f_res}b shows our
final rate measurements plotted against the intensity of the repumper laser. The observations
are consistent with the theoretical value for the slight dependence on 866 nm intensity.
Our final result for the lifetime was obtained with a slope fixed at the theoretical value
(see section \ref{s_tau}).

The radiation at 397 nm is obtained by frequency-doubling and so does not contain
any significant background light. There is no transition from $3D_{5/2}$ near enough to this
wavelength to cause significant de-shelving. A check for this or some other (unidentified)
effect was nevertheless carried out, where we changed the power of
this light by a factor 2, and we observed, to ten percent precision, no effect 
on the deshelving rate.

\subsubsection{Thermal radiation}

The rate of de-shelving is
$(1-b) B_{12} \rho(\omega_{21})$, where $\rho(\omega)$ is
the energy density per unit frequency interval in the thermal
radiation, and
$B_{12} = g_2 \pi^2 c^3 A_{21}/( g_1 \hbar
\omega_{21}^3 )$. In a thermal cavity we obtain the rate
  \beq
(1-b) \frac{2 J_2 + 1}{2 J_1 + 1} A_{21} e^{-\hbar \omega_{21}/k_{\rm B} T}
\label{thermal}
  \eeq
for $\hbar \omega_{21} \gg k_{\rm B} T$. For the 854~nm transition
in a room-temperature cavity, the rate is of order
$10^{-18}$~s$^{-1}$ so is negligible. Non-negligible rates
can be obtained, however, when we consider the radiation produced
by room lights or the filament of an ion gauge. The spectral
energy density is reduced
from the value in equation (\ref{thermal})
by a geometrical factor approximately equal to $S/(4 \pi r^2)$
where $S$ is the area of the hot filament and $r$ is its distance
from the ion, if the ion is within line of sight of the filament.
For example, taking $T=1700$ K, $S=20$ mm$^2$, $r=30$ cm, we
obtain a rate $\sim 5 \times 10^{-3}$~s$^{-1}$. In our vacuum system,
although the ion gauge is at this distance from the trapped ion,
it is not in line of sight, so we expect the rate of this process
to be well below this value.

\subsubsection{Collisional effects}

The ion in the shelved level can undergo a collision with an atom of the background
gas.  Either or both of two processes may then occur: the ion may gain a significant amount
of kinetic energy, and it may be transferred to another state.  This is a source of error since
in both cases the apparent shelving time --- the interval during which fluorescence is not
observed --- will be affected.  To investigate the nature and frequency of collisional effects we
monitored the fluorescence from the ion for 8 hours with the 397 nm and 866 nm radiation
present but with no laser operating at 850 nm to take the ion to the $4P_{3/2}$ level.  The
diffraction grating and aperture were in place.  During this period we observed abrupt
disappearance of the fluorescence (within the resolution of the 22 ms bins) on 17 occasions. 
It reappeared after times of the order of a second, with 6 ``dark periods" as short as a few
tens of milliseconds.  The reappearance was generally abrupt, but in 5 of the longer periods
it was more gradual, occurring over several bins.

       One non-collisional effect which can lead to loss of fluorescence in this test is
shelving in the $3D_{5/2}$ level.  Shelving caused by the 866 nm radiation
exciting the $3D_{3/2}$ to
$4P_{3/2}$ transition is negligible, because the spontaneous decay to $3D_{5/2}$ has
such a low branching
ratio (the excitation rate can be calculated from the branching ratio $b$ and the 
$3D_{3/2} \rightarrow 4P_{3/2}$ coefficient
$A_{21} = 0.91 \times 10^6$~s$^{-1}$ \cite{95:Liaw,89:Theodosiou}).
In contrast, it is not possible to rule out excitation by the spontaneously emitted light
from the 866 nm laser at 850 nm because the rate depends on the spectral distribution of the
spontaneous emission at 850 nm; if a peak happened to be close to the frequency of the 850
nm transition the process could be responsible for a significant number of the observed
events.  However, we would then expect an abrupt reappearance of the fluorescence on a
time scale of the order of a second, and the data then suggest that the upper limit for events
of this type is around eight.  These could alternatively be caused by fine-structure changing
collisions; the ion in the $3D_{3/2}$ level can be transferred to $3D_{5/2}$ by a long range collision
which may not transfer significant kinetic energy.  At our working pressure of order
$10^{-11}$ mbar we would expect
about 8 such events on the basis of an approximate value of the collisional mixing rate which
has been determined for conditions similar to ours \cite{93:Arbes,94:Arbes}. Our observations thus
provide a rough upper limit on the background gas pressure in our system
directly at the location of the ion, to confirm our ion gauge reading.
We note that the rates for 
$D_{5/2} \rightarrow D_{3/2}$ and $D_{3/2} \rightarrow D_{5/2}$ are approximately equal
at room temperature, therefore the present experiment does give an indication
of the de-shelving rate due to fine-structure changing collisions
in our $D_{5/2}$ lifetime measurements.

The other events are more obviously characteristic of collisions.  In particular, the
more gradual reappearance of the fluorescence after a long dark period is likely to be
associated with the ion being cooled again after acquiring a significant amount of kinetic
energy.  On the basis of our measured pressure and reasonable estimates of cross-sections,
we do not expect more than two or three collisions of this type. Some of the very short dark
periods are likely to be much smaller perturbations by relatively distant collisions.  If such a
perturbation were to occur during the lifetime measurement itself, fluorescence would not
be observable for a period whatever state the ion was in after the collision.  Fortunately, for
the purposes of estimating the uncertainty in the lifetime measurement, detailed interpretation
of the events is not necessary.  
An upper limit to the error introduced can be found by
assuming that all 17 events are due to collisions, giving a
rate of $6 \times 10^{-4}$ s$^{-1}$, and that such
a collision occurring in the experimental runs themselves while the ion was in the shelved
level would have delayed (or hastened)
the reappearance of fluorescence by an average of order 1 second. 
The net contribution to the measured rate has thus an upper limit
of $\pm 6 \times 10^{-4}$~s$^{-1}$
which is an order of magnitude below the statistical error.

\subsection{Systematic effects in the measurement process}

Both the fluorescence and the background signal on which it is superposed are subject to
fluctuations.  As well as the random error this introduces into the measurement, because the
instant at which fluorescence resumes is subject to statistical uncertainty, there are systematic
effects.  For example, a large fluctuation in the background can suggest that the ion has
decayed while it is still shelved; our data analysis procedure had to be developed and tested
to minimise errors due to such effects. Further, the ion is not cooled during the shelved
period, and if there is significant heating the fluorescence may be reduced for a time after
the decay.

\subsubsection{Heating during the shelved periods}  \label{s_heat}

In our preliminary experiments, we found that when an electric field was applied in the
vertical direction the fluorescence reappeared only gradually after shelving, over periods of
10--50 ms.  We ascribe this to heating. When the d.c.\ electric field
in the trap is not zero, the ion experiences the r.f.\
driving field, which is much more noisy than the d.c.\ field, and
so the ion motion heats during the shelved periods when it
is not laser cooled. We calculate that if the ion heats up
to room temperature,  it would take the lasers approximately
50~ms to cool the motion down again and thus for
the fluorescence to reappear. Evidence supporting this interpretation
is given by the fact that the non-abrupt
reappearance of fluorescence was correlated to the duration
of the dark period, being more likely for longer dark periods.
We therefore took care to ensure this phenomenon was not
present in the runs used for our final data set. This was
done by nulling the vertical field carefully before each run.

In our linear trap geometry the r.f.\ field is 2-dimensional.
To null the vertical field we adjusted the voltage on one
of the d.c.\ field
compensation electrodes, so as to minimise the linewidth
of the ion fluorescence as a function of 397~nm laser
frequency (at lowered 397 nm laser power).
The 397~nm laser beam used for shelving measurements
enters along the direction $(\sqrt{3},0,1)$ (where the $xz$ plane
is horizontal); this is sensitive to horizontal micromotion and hence
the vertical field.
We were able to null this field to $\pm 3$ V/m,
and most runs were carried out with nulling to $\pm 10 $ V/m.
We measured the
remaining heating rate, when this field was as well compensated
as we could make it, by blocking the cooling
laser beams (397~nm and 866~nm) for long periods, and looking
for a non-abrupt
return of the fluorescence when the beams were unblocked.
No delay was observed (the limit of sensitivity being the bin size
of 22 ms) unless
the cooling lasers were blocked for more than 10 minutes, some 600
times longer than the shelving times occurring during the measurements
themselves. The consequent systematic error in the lifetime can thus
be safely neglected; heating is likely to be relatively slow when
the ion is first shelved, but even assuming that the delay of
fluorescence is linear with shelving time the error is only of order
30~$\mu$s. We note that there were no soft edges in the runs used
for our final data set.

We found that a horizontal field gave a much smaller heating
effect, and indeed some of our final data points were taken in
the presence of a horizontal field of order 200 V/m.
However, for our final experimental run,
a further beam was introduced along $(0,-2,-1)$ to allow
horizontal field compensation; this beam was blocked
once the compensation was optimised. Although not important for
the present lifetime measurements, accurate compensation is
necessary for our future work in which the ion is to be cooled
below the Doppler limit.

\subsubsection{Tests of the reliability of the data analysis}

We consider three possible sources of systematic error in the data analysis in turn.  
First, it is straightforward to show analytically that approximating the continuous 
exponential distribution of dark times as a histogram causes no systematic errors in 
the fitted value of \( \tau \).  Second, numerically simulated data was used to show 
that varying the threshold level used in the data analysis, did not systematically 
change the fitted value of \( \tau \) by more than $0.1$ ms.  
Our real data has a further property: intensity
fluctuations. The $\tau$ value deduced from a real data set was found to
vary as a function of threshold by amounts of order $\pm 0.5$ ms. We estimate
this may introduce a systematic error of order $0.5$ ms.
Third, repeated fits to 
numerically simulated data sets were used to show that any systematic error arising
purely from the exponential fitting procedure was less than $0.25$ ms.  They also
permitted us to verify the uncertainty (one standard deviation)
in the fitted value of $\tau$, derived from
the cost function surface.

These effects give a total uncertainty of $\pm 0.6$ ms.

\section{Demonstration of photon antibunching} \label{s_anti}

\begin{figure}[h]
\centerline{\resizebox{15cm}{!}{\includegraphics{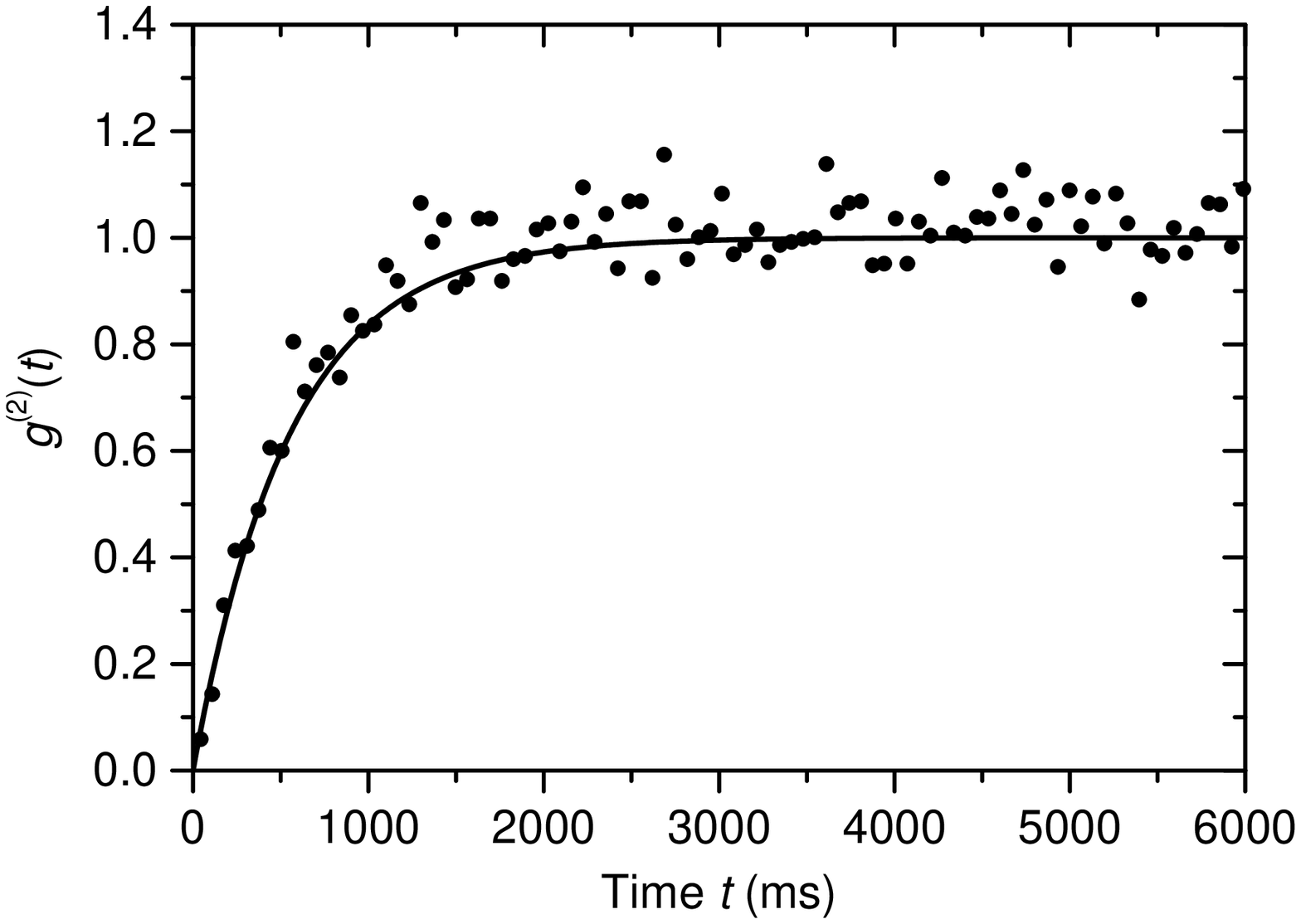}}}
\rule{0pt}{24 pt}
\caption{Second-order coherence of the 729 nm radiation emitted by the
ion, as deduced from quantum jump observations. The data points give
the number of times a jump was observed (therefore a 729 nm photon emitted)
at time $t$ after a chosen jump, normalised to the mean jump rate, accumulated
in several data sets. The line
is the theoretical expectation, described in the text. }
\label{f_g2}
\end{figure}

The quantum jump method provides a convenient means to demonstrate
photon antibunching \cite{Bk:Loudon}, that is, generation of a light field
whose second order coherence $g^{(2)}(t)$ falls below 1 as
$t \rightarrow 0$.
The second-order coherence is the normalised
autocorrelation function of the
intensity. A classical definition of intensity leads to
$g^{(2)}(t) < g^{(2)}(0)$ and
$g^{(2)}(0) > 1$, whereas a definition in terms
of quantum electric field operators allows all values \cite{Bk:Loudon}.
Therefore observation of a $g^{(2)}(0)$
value below 1 is a direct signature of the quantum nature of the
radiation field. Quantum jump observations yield $g^{(2)}$
easily, since whenever a de-shelving jump is observed, we may
deduce, with close to 100\% reliability,
that one photon at 729~nm has been emitted by the ion
\cite{86:Bergquist}. Therefore
we deduce the presence of a 729~nm radiation field
which has $g^{(2)}(t) = \left< n(t) n(0) \right>/\bar{n}^2$,
where $n(t)$ and $n(0)$ are the numbers of jumps observed in two
equal time intervals separated by a time $t$. This quantity
is plotted in figure \ref{f_g2} for a typical data set in our experiment.
For this study, we did not use the method described in section
\ref{s_method}, in which the 850~nm light was blocked when
the ion was shelved. Instead, this light was left permanently
on, and the fluorescence monitored continuously. We therefore
obtained the well-known `random telegraph' signal, figure \ref{f_sig}b.
Since we detect very close to all the 729~nm photons, 
we can be confident the
second order coherence does not vary at time-scales too short for
us to detect. Therefore the antibunching
signature $g^{(2)}(0) < 1$ is clearly
demonstrated. 

We can understand the complete form of $g^{(2)}(t)$ by solving
the rate equations for the populations of the ion's levels. All the
relevant processes are fast except for two, namely spontaneous
decay from $D_{5/2}$ to $\Lambda$ and excitation from
$\Lambda$ to $D_{5/2}$, so the problem reduces to a two-level
system. If we take $t=0$ to be the centre of a short time interval
(one bin) during which a jump occurred, so that $n(0)=1$, then
the probability for the atom to emit a 729~nm photon
at time $t$ is proportional to the population of $D_{5/2}$
at $t$, given that it was zero at $t=0$. We thus obtain
  \beq
g^{(2)}(t) = 1 - e^{-(R+\gamma) t}
  \eeq
where $R$ is the excitation rate, $\gamma$ the decay rate. This
curve is plotted on figure \ref{f_g2} with no free parameters ($R$ is
taken as the inverse of the mean duration of bright periods in the
same data set). The agreement between data and
theory is evidence that we have a good understanding
of our experimental situation.

\section{Result and discussion}   \label{s_tau}

Our final data set consisted of four 8-hour runs, and two 2-hour
runs, all using the experimental method of section \ref{s_method}
(850~nm laser blocked during dark periods). The de-shelving rates
observed in these runs, as obtained by the analysis described above,
are shown in figure \ref{f_res}b. The size of each error bar is equal
to the statistical uncertainty emerging from the analysis. The straight
line through the data is a single-parameter weighted
least-squares fit. The line
is given the theoretically expected slope, and the
best fit intercept is found to be $\gamma_0 = 0.856 \pm 0.004$~s$^{-1}$.

The de-shelving rate $\gamma$ obtained in the experiment may be written
  \beq
\gamma = \frac{1}{\tau} + \sum_i \gamma_i
  \eeq
where $\tau$ is the natural lifetime we wish to measure, and
the $\gamma_i$ are due to other processes
which contribute to the measured rate, chiefly laser excitation of transitions in 
the ion and collisional processes. The effect of off-resonant excitation by 866nm radiation 
is already accounted for in our fitted intercept $\gamma_0$. However, uncertainty ($\pm 30\%$) 
in the light intensity at the ion makes this accounting imprecise, leading to a further 
systematic error in $\gamma_0$ at the level of 0.001\,s$^{-1}$. This and other contributions 
to the systematic error, from collisional processes and remaining background light from the
866nm laser, are shown in Table~\ref{t_sys}. The statistical error from the data 
fitting dominates. The systematic effects are independent of one another, so we add their
errors in quadrature. Adding this total systematic error linearly to the statistical
error, the value we obtain for the natural lifetime of the $D_{5/2}$ level is
\beq
\tau = 1.168 \pm 0.007 \mbox{ s}
\eeq

\begin{table}
\small
\begin{tabular}{lll}
 effect		& \multicolumn{2}{c}{$1\sigma$ errors ($10^{-3}$\,s$^{-1}$)} \\
					&   statistical	&	systematic \\
\hline
data fitting				&	$\pm 4$	&	$\pm 0.4$		\\
866nm intensity uncertainty (30\%)	&			&	$\pm 1$		\\
$A_{21}$ uncertainty (4\%)              &                       &       $\pm 0.1$     \\
collisional processes			&			&	$\pm 0.6$	\\  
background from 866nm laser		&			&	$\sim + 0.01$	\\
\hline
total					&     $\pm 4$		& $\pm 1.2$
\end{tabular}
\caption{
Statistical and systematic contributions to the error in the measured de-shelving
rate $\gamma$. The systematic errors are added in quadrature, then their total
is added linearly to the statistical error, to 
give a total uncertainty of $5.2\times 10^{-3}$\,s$^{-1}$. See also
section~\protect\ref{s_sys}.}
\label{t_sys}
\end{table}

We have already noted the significant difference between this result and
earlier measurements. Here we compare it with theory. Our result differs
from all the reported {\em ab initio}
calculations by amounts large compared with
our $0.6$ \% experimental uncertainty.
The relative difference $(\tau_{\rm theor.} - \tau_{\rm meas.})/\tau_{\rm meas.}$
is $-10$\% for a recent calculation
based on the Brueckner approximation \cite{95:Liaw}, 
$+5.6$\% for relativistic many-body perturbation theory \cite{91:Guet},
and $-2.6$\% for multi-configurational Hartree Fock
(MCHF) calculations \cite{92:Vaeck,93:Brage}.
Of these calculations, those 
which produced the smallest discrepancy with our measured $D$ state lifetime
produced the largest discrepancy with the measured $P$ state lifetimes
obtained by Jin and Church \cite{93:Jin}.
A semi-empirical calculation
based on MCHF and core polarization \cite{92:Vaeck}, and a 
calculation using related methods \cite{93:Brage}, 
give a value close to our measured result (relative difference $-0.7$\%).
The most natural interpretation of these observations is that the {\em ab initio}
calculations of $\tau$ carried out so far have a precision 
of at best a few percent, and success at calculating electric dipole matrix elements
does not guarantee the same degree of success with other parameters such
as electric quadrupole matrix elements.

The ratio between the lifetimes of $D_{3/2}$ and $D_{5/2}$ is much less sensitive
to imperfections in the calculations, and is given primarily by the
frequency factor $\omega^5$, which leads to a ratio $1.022$. The factor
emerging in recent calculations of the two lifetimes is
$1.0335$ \cite{95:Liaw}, $1.0283$ \cite{91:Guet}, $1.0175$
\cite{92:Vaeck,93:Brage}. We take the standard deviation of these results
to indicate the theoretical uncertainty on their mean, giving
$\tau_{3/2} / \tau_{5/2} = 1.026 \pm 0.007$. Combining this with our
measurement of $\tau_{5/2}$ gives $\tau_{3/2} = 1.20 \pm 0.01$ s
for the lifetime of the $3d\,^2\! D_{3/2}$ level in $^{40}$Ca$^+$.

\section{Conclusion}

We have described a new linear ion trap apparatus and its use
to measure the lifetime of the $D_{5/2}$ level in $^{40}$Ca$^+$ by
quantum jump measurements on a single trapped ion. 
Our result
is more precise than previous measurements, and significantly
larger. We believe this discrepancy is mostly explained by
previously unrecognised systematic errors, which tend to
make the lifetime appear shorter than it in fact is. 
Our measurement provides a new precise test of {\em ab initio}
atomic structure calculations; it is in only moderate
agreement with currently reported calculations. 

We have discussed the spectral distribution of light emitted
by a semiconductor diode laser operated in an extended cavity,
emphasizing that in experiments where unwanted
excitation of allowed atomic transitions is to be avoided,
it is necessary to take into account the weak
broad component in the spectrum, which extends many nanometres
away from the main lasing wavelength. This is significant
in many experiments in atomic physics, especially coherent
atom optics, `dark' optical lattices, frequency standards
and quantum information processing.

Finally, we have discussed various diagnostics on the performance
of the ion trap, made possible by careful analysis of long
periods of operation of the trap. These include an upper limit for
the pressure in the system, and the heating rate in the trap.
We demonstrated photon anti-bunching, finding it
in agreement with theoretical expectations.

Our uncertainty is dominated by statistics, so it would be possible
to obtain significantly higher precision by accumulating more data,
though for single-ion work
this would require integration for several weeks. To go further,
one would need better knowledge of the laser
intensities and collisional processes.

\section{Acknowledgements}

We would like to acknowledge helpful correspondence with G. Werth, and
a preprint of \cite{99:Block}. 
L. Favre, E. Hodby, M. McDonnell and J. P. Stacey contributed to
development of the apparatus, D. T. Smith designed the high voltage r.f. supply,
and we are grateful for general technical assistance from G. Quelch.
We thank R. Blatt and the Innsbruck group
(especially  F. Schmidt-Kaler, C. Roos, M. Schulz)
for many helpful discussions
and guidance, and similarly R. C. Thompson and D. N. Segal.
This work was supported by EPSRC (GR/L95373), the Royal Society,
Riverland Starlab and Oxford University (B(RE) 9772).

\newpage
\onecolumn

\newpage
%\rule{2 cm}{0 cm}
%\newpage

\newpage

\newpage

\newpage

\newpage

\end{document}